\documentstyle[12pt]{article}

\def\journal{\topmargin .3in    \oddsidemargin .5in
        \headheight 0pt \headsep 0pt
        \textwidth 5.625in 
\textheight 8.25in 
        \marginparwidth 1.5in
        \parindent 2em
        \parskip .5ex plus .1ex         \jot = 1.5ex}
\journal

\input{epsf}
\begin{document}
\begin{titlepage}
\begin{center}
July 17, 1998      \hfill    LBNL-42103\\

\vskip .5in

{\large \bf Higgs boson 
mass constraints from precision data and direct searches}\footnote
{This work was supported by the Director, Office of Energy
Research, Office of High Energy and Nuclear Physics, Division of High
Energy Physics of the U.S. Department of Energy under Contract
DE-AC03-76SF00098.}

\vskip .5in

Michael S. Chanowitz\footnote{Email: chanowitz@lbl.gov}

\vskip .2in
{\em Theoretical Physics Group\\
     Lawrence Berkeley National Laboratory\\
     University of California\\
     Berkeley, California 94720}
\end{center}
\vskip .25in 

\begin{abstract} 

Two of the nine measurements of $sin^{2}\theta^{lepton}_{eff}$, the 
effective weak interaction mixing angle, are found to be in 
significant conflict with the direct search limits for the Standard 
Model (SM) Higgs boson.  Using a scale factor method, analogous to one 
used by the Particle Data Group, we assess the possible effect of 
these discrepancies on the SM fit of the Higgs boson mass.  The scale 
factor fits increase the value of $sin^{2}\theta^{lepton}_{eff}$ by as 
much as two standard deviations.  The central value of the Higgs boson 
mass increases as much as a factor of two, to $\simeq 200$ GeV, and 
the 95\% confidence level upper limit increases to as much as 750 GeV. 
The scale factor is based not simply on the discrepant measurements, 
as was the case in a previous analysis, but on an aggregate 
goodness-of-fit confidence level for the nine measurements and the 
limit.  The method is generally applicable to fits in which one or 
more of a collection of measurements are in conflict with a physical 
boundary or limit. In the present context, the results suggest 
caution in drawing conclusions about the Higgs boson mass from the 
existing data.

\end{abstract} 
\end{titlepage}

\renewcommand{\thepage}{\roman{page}}
\setcounter{page}{2}
\mbox{ }

\vskip 1in

\begin{center}
{\bf Disclaimer}
\end{center}

\vskip .2in

\begin{scriptsize}
\begin{quotation}
This document was prepared as an account of work sponsored by the United
States Government. While this document is believed to contain correct
 information, neither the United States Government nor any agency
thereof, nor The Regents of the University of California, nor any of their
employees, makes any warranty, express or implied, or assumes any legal
liability or responsibility for the accuracy, completeness, or usefulness
of any information, apparatus, product, or process disclosed, or represents
that its use would not infringe privately owned rights.  Reference herein
to any specific commercial products process, or service by its trade name,
trademark, manufacturer, or otherwise, does not necessarily constitute or
imply its endorsement, recommendation, or favoring by the United States
Government or any agency thereof, or The Regents of the University of
California.  The views and opinions of authors expressed herein do not
necessarily state or reflect those of the United States Government or any
agency thereof, or The Regents of the University of California.
\end{quotation}
\end{scriptsize}

\vskip 2in

\begin{center}
\begin{small}
{\it Lawrence Berkeley National Laboratory is an equal opportunity employer.}
\end{small}
\end{center}

\newpage

\renewcommand{\thepage}{\arabic{page}}
\setcounter{page}{1}

\noindent{\bf {1. Introduction}} 

Beautiful measurements of $Z$ boson decay asymmetries at LEP and 
SLC\cite{ewwg98} and of the top quark mass at Fermilab\cite{98mt} 
appear to constrain the mass of the Standard Model (SM) Higgs boson at 
the level of a factor two or better.  The combined fit of nine 
measurements of the effective leptonic weak interaction mixing angle 
yields $sin^{2}\theta^{lepton}_{eff}=0.23148\pm 0.00021$, which 
implies the SM Higgs boson mass $m_{H}=86^{+84}_{-42}$ GeV and the 
upper limit $m_{H}< 260$ GeV at 95\%
confidence level (CL).  In a previous letter\cite{msc} I observed that 
the most precise of the nine measurements,  
the left-right asymmetry $A_{LR}$, then implied $m_{H}=16$ 
GeV and an {\em upper} limit $m_{H}<77$ GeV at 95\%CL,
in contrast to the {\em lower} limit from direct searches, then given 
by $m_{H}>77$ GeV, also at 95\%CL. I analyzed the possible impact of 
this discrepancy on the SM fit of $m_{H}$ using a scale factor method 
inspired by a method the Particle Data Group\cite{pdg} (PDG) has used 
to combine discrepant data.  The conclusion was that both the central 
value and the upper limit on $m_{H}$ could be appreciably higher than 
in the conventional fit.  Similar observations had been made 
previously, using different methods, by Gurtu\cite{gurtu} and 
Dittmaier, Schildknecht, and Weiglein.\cite{dsw}

The work presented here differs significantly from reference 
\cite{msc} in which the discrepancy between the $A_{LR}$ measurement 
and the search limit was evaluated simply as the likelihood for a 
95\%CL upper limit at 77 GeV to be consistent
with a 95\%CL lower limit at the same mass, i.e., $2\cdot0.05\cdot0.95
\simeq 0.1$ or 10\%.  This may be a fair appraisal if we have an {\em 
a priori} reason to focus on the $A_{LR}$ measurement, such as for 
instance that it provides the most precise determination of 
$sin^{2}\theta^{lepton}_{eff}$, rather than choosing to consider it 
{\em because} we have noticed that it implies a value of $m_{H}$ below 
the SM search limit.  In the latter case we need to consider the 
likelihood that {\em any} of the nine relevant measurements of 
$sin^{2}\theta^{lepton}_{eff}$ could fluctuate to produce a like 
discrepancy.  It is fair to say that in this instance our attention is 
drawn to $A_{LR}$ by both its precision and the fact of its conflict 
with the SM search limits.

It may therefore be appropriate to approach the analysis from the 
perspective of the consistency of the complete ensemble of nine 
measurements with the SM search limit.  That is the perspective of the 
analysis presented here, in which a suitable scale factor method is 
proposed.  The method can be applied to a variety of different 
physical situations, for instance, the problem confronted by the PDG 
of how to set an upper limit on the electron neutrino mass when 
several measurements (of the kinematic end-point in tritium decay) 
imply a tachyonic mass.\cite{pdg} Here I will apply the 
method to the SM fit of $m_{H}$, using the Spring 1998 data, which 
differs appreciably from the Summer 1997 data used in the earlier 
analysis.

In the previous analysis the scale factor was introduced based on the 
goodness-of-fit CL between just the discrepant measurement and the 
limit.  In the method presented here the scale factor is determined by 
the goodness-of-fit CL between the complete set of asymmetry 
measurements and the limit, therefore taking account of the likelihood 
that any measurement in the set might fluctuate into the low 
tail of the $sin^{2}\theta^{lepton}_{eff}$ distribution.  The method 
is then truly analogous to the PDG method, which rescales the fit 
uncertainty by a scale factor determined by the goodness-of-fit CL of 
the chi-squared distribution of the complete data set.

It is important to keep in mind that the analysis presented here 
assumes the validity of the Standard Model (or the MSSM in the 
decoupling limit) and that in general, without a specific theoretical 
framework, the electroweak radiative corrections tell us nothing about 
the nature of electroweak symmetry breaking.  In addition to quantum 
corrections from the Higgs sector, the value of 
$sin^{2}\theta^{lepton}_{eff}$ could be affected by quantum 
corrections from other sectors of new physics and/or from gauge boson 
mixing in theories with extended gauge sectors.  The nature of 
electroweak symmetry breaking can only be definitively established by 
direct discovery and detailed study of the Higgs sector quanta at a 
high energy collider.  Until then anything is possible: light Higgs 
scalars, dynamical symmetry breaking without Higgs scalars, or even 
that the Higgs mechanism is not realized in nature at all.  Here we 
assume that no new physics contributes to 
$sin^{2}\theta^{lepton}_{eff}$ except the quantum corrections from the 
Higgs sector, and that any Higgs scalar decays as prescribed in the SM 
so that the Higgs boson search limits are applicable.

Section 2 is a brief review of the 1998 data and the SM fit of 
$m_{H}$.  The uncertainties in the fit are examined for two different 
evaluations of $\alpha(m_{Z})$.\cite{ej,dh98} (The values quoted 
in this introductory section are based on reference \cite{ej}.)  
Though the 1998 data set for $sin^{2}\theta^{lepton}_{eff}$ is more 
internally consistent than the 1997 data, its confidence level is 
still not robust and it continues to exhibit discrepancies with the SM 
search limits.  The central value of $m_{H}$ implied by $A_{LR}$ has 
increased to 25 GeV, but the direct search limit\cite{hsearches} has 
also increased, to $m_{H}>89.3$ GeV at 95\%CL, and the
precision of the $A_{LR}$ measurement has improved.  Putting all these 
changes together there is still a significant discrepancy, with 
$A_{LR}$ now implying $m_{H}<89.3$ GeV at 93\%CL.

A somewhat bigger discrepancy occurs in the less precise tau 
front-back asymmetry measurement, $A_{FB}^{\tau}$, which implies 
$m_{H}=4$ GeV and $m_{H}<89.3$ GeV at 95\%CL. Although a single value
of $sin^{2}\theta^{lepton}_{eff}$ is typically presented for the 
combined leptonic front-back asymmetry, $A_{FB}^{l}$, the measurements 
of $A_{FB}^{e}$, $A_{FB}^{\mu}$, and $A_{FB}^{\tau}$ are in fact quite 
distinct, each posing a unique set of experimental issues.  As can be 
seen in table 2 below, $A_{FB}^{\mu}$, and $A_{FB}^{\tau}$ are 
individually at the same level of precision as all but the two most 
precise measurements, so it is most natural to consider them 
separately.  

It is certainly the case that our attention is drawn to 
$A_{FB}^{\tau}$ by the low value of $m_{H}$ it implies, so in 
considering the conflict of $A_{LR}$ and $A_{FB}^{\tau}$ with the 
search limit we must assess the goodness-of-fit of the measurements 
with the search limit from the perspective of the complete set of nine 
measurements.  The scale factors computed in this way then 
appropriately weight the increased likelihood of outlying measurements 
when $A_{FB}^{l}$ is disaggregated, with the number of 
$sin^{2}\theta^{lepton}_{eff}$ measurements increased from seven to 
nine.

Section 3 begins with a review of the PDG scale factor method for 
combining discrepant data and then presents a method to extend it to 
the case of measurements in conflict with a limit.  The central 
observation of the PDG is that low CL data sets occur more often than 
expected by chance, and that historically many discrepancies are found 
to result from underestimated systematic errors.  This should not be a 
surprise, since the estimation of systematic error is perhaps the most 
challenging task faced by experimenters in the analysis and 
presentation of their data.  The PDG scaled error is meant to provide 
a more cautious interpretation of low CL data sets, with minimal 
impact on moderately discrepant data.  After reviewing the motivation 
and formulation of the PDG scale factor, $S^{*}$, an analogous scale 
factor is constructed for situations in which the discrepancy is 
between a collection of measurements and a limit.  Section 3 concludes 
with a brief discussion of the complementary relationship of the scale 
factor method with a recent analysis by Cousins and Feldman\cite{cf} 
of confidence intervals near a physical boundary.  Their construction 
is used to determine the upper limits on $m_{H}$ from the scaled fits.

Section 4 presents the application of the scale factor method to the 
fit of $m_{H}$ from the nine measurements of 
$sin^{2}\theta^{lepton}_{eff}$.  The result is a continuum of fits 
which differ in how the scaling is shared between the two low 
measurements, $A_{LR}$ and $A_{FB}^{\tau}$.  At one extreme, it 
suffices to scale the uncertainty of $A_{FB}^{\tau}$ by a factor 3 
while leaving $A_{LR}$ unmodified; in this case the effect on the fit 
is small.  At the other extreme, when the rescaling is dominantly 
applied to $A_{LR}$, the fitted central value of $m_{H}$ 
increases by a factor two relative to the conventional fit, while the 
95\%CL upper limit (in the Cousins-Feldman
construction) increases by nearly a factor three relative to the 
conventional 95\%CL limit.  These extremes and
a sample of intermediate cases are presented in Section 4.

The analysis in sections 2 -- 4 assumes a perfect search limit, 
$m_{H}>89.3$ GeV with 100\% CL. In section 5 I show that the results 
obtained in this approximation apply to the actual, less than perfect 
experimental limits.  The conclusion relies on the sharply increased 
confidence level obtained by the search experiments for values of 
$m_{H}^{\rm LIMIT}$ slightly below 89 GeV.

A brief summary and discussion are given in Section 6.

\noindent{\bf {2. The electroweak data and the SM Higgs boson mass}}

Our strategy is to focus on the most direct determination of $m_{H}$, 
using the measurement of $sin^{2}\theta^{lepton}_{eff}$, augmented by 
the direct measurement of the top quark mass (by CDF and D0) together 
with the value of $\alpha(m_{Z})$.  The effective mixing angle, 
$sin^{2}\theta^{lepton}_{eff}$, has the greatest sensitivity to 
$m_{H}$ with the least collateral dependence on various other 
quantities such as the strong coupling constant $\alpha_{S}(m_{Z})$ or 
the fraction of hadronic $Z$ decays to $b$ quarks, $R_{b}$.  From the 
nine measurements of $sin^{2}\theta^{lepton}_{eff}$, which combine to 
yield $sin^{2}\theta^{lepton}_{eff}=0.23148\pm0.00021$, and the 
conservative determination of $\alpha(m_{Z})=(128.896 \pm 0.090)^{-1}$ 
by Eidelmann and Jegerlehner\cite{ej} I obtain using the state of the 
art radiative corrections of Degrassi et al.\cite {dgps} 
$m_{H}=86^{+84}_{-42}$ GeV, compared with the EWWG\cite{ewwg98} global 
fit value $m_{H}=66^{+74}_{-39}$ GeV (which also uses reference 
\cite{ej} for $\alpha(m_{Z})$).  Gaussian statistics are assumed for 
the $sin^{2}\theta^{lepton}_{eff}$ measurements, from which it follows 
in the SM fit that the logarithm of the Higgs boson mass, ${\rm ln}\ 
m_{H}$, is Gaussian distributed.

The difference between the global fit and the fit based just on the 
$sin^{2}\theta^{lepton}_{eff}$ data is not great and is due primarily 
to the fact that the global fit uses the top quark mass, $m_{t}=171.1 
\pm 5.1$ GeV, determined from the combination of direct and indirect 
measurements, while in the fit restricted to the 
$sin^{2}\theta^{lepton}_{eff}$ data I have used the directly measured 
Fermilab value\cite{pdg}, $m_{t}=173.8 \pm 5.1$ GeV. The smaller value 
of $m_{t}$ from the indirect determination is due principally to the 
remnant of the $R_{b}$ anomaly --- since the current value of $R_{b}$ 
is 1.6 standard deviations above the SM fit value, the global fit 
prefers smaller values of $m_{t}$ in order to minimize the 
discrepancy.  Because $m_{t}$ and $m_{H}$ are correlated in the fit, a 
higher value of $R_{b}$ thus leads indirectly to a lower value of 
$m_{H}$ in the global fit.  Since in this paper I am assuming the 
validity of the Standard Model, the strategy followed seeks to 
minimize the extent of such indirect effects, which during the height 
of the $R_{b}$ anomaly (when $R_{b}$ was believed to be three standard 
deviations above the SM value) led to a serious distortion of the 
global fit of $m_{H}$.\cite{dsw}

The uncertainty in the SM determination of $m_{H}$ is analyzed in 
table 1.  The principal sources of uncertainty are the uncertainties 
in the measurements of $sin^{2}\theta^{lepton}_{eff}$ and $m_{t}$, and 
the evaluation of the fine structure constant at $m_{Z}$.  I use 
$sin^{2}\theta^{lepton}_{eff}= 0.23148 \pm 0.00021$ from the 
conventional least square fit of the nine measurements and $m_{t} = 
173.8\pm 5.1$ GeV from the current PDG fit of the Fermilab top quark 
mass measurements.  For $\alpha(m_{Z})$ I use two values, $(128.896 
\pm 0.090)^{-1}$ and $(128.933 \pm 0.021)^{-1}$.  The former is the 
conservative evaluation by Eidelmann and Jegerlehner\cite{ej}, while 
the latter, from Davier and H\"ocker\cite{dh98}, is one of 
several\cite{otheralpha} recent, more optimistic evaluations, which 
rely on perturbative QCD down to lower energy scales.  These typically 
have a smaller estimated error and a smaller central value, the latter 
implying a larger value of $m_{H}$.  In this paper I will present 
results using both references \cite{ej} and \cite{dh98}.  Table 1 also 
displays much smaller contributions from the QCD coupling constant, 
$\alpha_{S}(m_{Z})=0.120 \pm 0.003$, and from uncomputed higher order 
corrections.  For the latter I rely on the estimate of Degrassi et 
al.\cite{dgps}, whose compact representation of their calculations of 
the radiative corrections are used throughout this 
paper.\footnote{Weiglein and coworkers\cite{weiglein} have recently 
estimated a somewhat larger theoretical error for the results of 
reference \cite{dgps}.  However in any case the theoretical error is 
overwhelmed by the three dominant uncertainties in table 1.} Combined 
in quadrature the net uncertainty in $\rm{ln}(m_{H})$ is $\pm 0.67$ or 
$\pm 0.52$ for the two evaluations of $\alpha(m_{Z})$, corresponding 
respectively to a factor 2 or 1.7 uncertainty in $m_{H}$.

The measurements of $sin^{2}\theta^{lepton}_{eff}$ have been 
characterized by three discrepancies, which persevere, though at a 
diminished level, in the Spring 1998 data.  In the Summer 1997 data 
the two most precise measurements, $A_{LR}$ and $A_{FB}^{b}$, differed 
by $3.1\sigma$ (CL = 0.002), and $A_{LR}$ differed from the LEP 
average by $2.9\sigma$ (CL = 0.005).  In the Spring 1998 data 
$sin^{2}\theta^{lepton}_{eff}$ from $A_{LR}$ has increased by 
$0.7\sigma$ while $sin^{2}\theta^{lepton}_{eff}$ from $A_{FB}^{b}$ has 
decreased by $0.6\sigma$, so that the corresponding discrepancies are 
$2.3\sigma$ (CL = 0.02) and $2.4\sigma$ (CL = 0.015).  The chi-squared 
for the nine measurements has improved from $\chi^{2}/dof=14.5/8$ (CL 
= 0.07) to a more acceptable $\chi^{2}/dof=10.7/8$ (CL = 0.2).  The 
nine measurements are shown in table 2 along with their ``pulls'', 
defined as the number of standard deviations that each measurement 
differs from the least-squares fit value $0.23148 \pm 0.00021$.  As 
another estimator of the consistency of the nine measurements I have 
used a Monte Carlo to compute the confidence level to replicate the 
observed distribution of the {\em absolute values} of the pulls, 
obtaining a probability of 0.07.\footnote{That is, 0.07 is the 
probability that the absolute value of the largest pull is $\ge 1.61$, 
the second $\ge 1.57, \ldots$ , and the ninth $\ge 0.01$.}

Tables 3 and 4 (corresponding to $\alpha(m_{Z})$ from references 
\cite{ej} and \cite{dh98} respectively) shows the Higgs boson mass 
predictions of each of the nine $sin^{2}\theta^{lepton}_{eff}$ 
measurements listed in order of precision. 
For each measurement the tables  display the  central 
value for $m_{H}$, the symmetric (in ln$(m_{H}))$ 90\% confidence 
interval, and the implied probability that $m_{H}$ lies {\em below} 
89.3 GeV, which is the current 95\%CL lower limit from the LEP direct 
searches.\cite{hsearches}.  To compute the confidence intervals in 
ln$(m_{H})$ and the implied probabilities for $m_{H}<89.3$ GeV we must 
of course include the parametric errors shown in table 1, for 
instance, by treating ln$(m_{H})$ as a Gaussian statistical variable 
for each measurement, combining in quadrature the uncertainty arising 
from the particular measurement of $sin^{2}\theta^{lepton}_{eff}$ with 
the other parametric errors shown in table 1.  Equivalently, as a 
matter of convenience, one may express the parametric errors as 
effective errors in $sin^{2}\theta^{lepton}_{eff}$ (e.g., for fixed, 
known $m_{H}$) and combine them in quadrature with the experimental 
$\delta(sin^{2}\theta^{lepton}_{eff})$.\footnote{ The theoretical 
uncertainties of the very large and very small values of $m_{H}$ in 
tables 3 and 4 are somewhat bigger than indicated in table 1.  The 
largest values, $\gg 1$ TeV, have no precise meaning in any case.  For 
the very small values, such as $m_{H}=4$ MeV from $A_{FB}^{\tau}$, we 
are really only concerned with the implied probability $P(m_{H}<89.3\ 
{\rm GeV})$ which only depends on the relationship between $m_{H}$ and 
$sin^{2}\theta^{lepton}_{eff}$ at $m_{H}=m_{H}^{\rm LIMIT}$ where 
table 1 does apply.}

The question we wish to consider is whether/how the discrepancies of 
$A_{LR}$ and $A_{FB}^{\tau}$ with the SM Higgs boson search limits 
should affect the SM fit of the Higgs boson mass.  The first 
part of the question is how big in fact is the discrepancy?  
The answer depends on precisely how we frame the question.  If, 
without considering the particular central value obtained, we had an 
{\em a priori} reason to focus on a particular measurement, say on 
$A_{LR}$ because it is the most precise and therefore most important 
single measurement in the fit, then the discrepancy could be read off 
from table 3 or 4 (though also including the effect of the less than 
perfect 95\% confidence level of the search
limit) and the analysis might then proceed as in reference \cite{msc}.  
However it is fair to say that in the present context our attention is 
drawn to $A_{LR}$ and $A_{FB}^{\tau}$ by the fact of their conflict 
with the search limits.  In that case the appropriately framed 
question is how likely is it that any two of the nine measurements 
could fluctuate to provide discrepancies with the search limits equal 
or greater than the observed discrepancies?  We obtain an upper limit 
on that probability by assuming that the true value of $m_{H}$ is 
precisely at the value of the direct search lower limit, $m_{H}=89.3$ 
GeV.

Let $p_{\tau}$ and $p_{LR}$ be the probabilities implied by the 
measurements of $A_{LR}$ and $A_{FB}^{\tau}$ that $m_{H}$ lies below 
89.3 GeV. Then the upper limit on the probability that any two of nine 
measurements, $a$ and $b$, could fluctuate into the low tail of the 
$sin^{2}\theta^{lepton}_{eff}$ distribution such that $p_{a}\geq 
p_{\tau}$ and $p_{b}\geq p_{LR}$ is given by\footnote{That is, 
$P_{9}(p_{\tau},p_{LR})$ is the complement of the probability that 
all nine measurements have $p_{i}< p_{\tau}$  or that one among them 
has $p_{i}>p_{\tau}$ while the other eight have $p_{i}< p_{LR}$. }
$$
P_{9}(p_{\tau},p_{LR})= 1 - p_{\tau}^{9} - 9(1 - 
p_{\tau})p_{LR}^{8}. \eqno{(1)}
$$
Equation (1) is the goodness-of-fit CL between the nine measurements 
and the direct search limit in the Standard Model, assuming the search 
limits to be perfect.  Taking $p_{\tau}$ and $p_{LR}$ from tables 3 
and 4 we find $P_{9}(p_{\tau},p_{LR})= 0.12$ and 0.18 respectively.  
Though we assume here that the search limit has 
100\%CL, it is shown in section 5 that essentially 
the same results are obtained when the actual confidence levels of the 
searches are taken into account.

These confidence levels, 0.12 and 0.18, might be characterized as 
marginal, not big enough to be considered ``robust'' nor small enough 
to force us to choose between the Standard Model and the experiments.  
They are in the gray area to which the Particle Data Group scaling 
factor $S^{*}$ would apply if similar CL's were obtained from the 
$\chi^{2}$ distribution of a collection of measurements, as 
discussed in the next section.
 
\noindent{\bf {3. Scale factors for discrepant data}}

Having quantified the extent of the discrepancy between the search 
limit and the measurements of $sin^{2}\theta^{lepton}_{eff}$ in the 
SM, we now consider the more difficult aspect of the question: 
whether/how these discrepancies should affect the SM fit of the Higgs 
boson mass.  There is no single ``right'' answer.  A 
maximum likelihood fit including both the precision data and the 
direct search data would replicate the conventional fit if the central 
value lies above the lower limit, $m_H^{\rm LIMIT}$, from the direct 
searches.  That is a defensible interpretation, since if the true 
value of $m_H$ were near $m_H^{\rm LIMIT}$ we would expect values of 
$m_H$ obtained from measurements of $sin^{2}\theta^{lepton}_{eff}$ to 
lie both above and below $m_H^{\rm LIMIT}$.  By underweighting 
downward fluctuations while leaving upward fluctuations at their full 
weight, we risk skewing the fit upward.  Mindful of this risk, it is 
still instructive to explore the sensitivity of the fit to the weight 
ascribed to measurements that are in significant contradiction with 
the direct search limit.

Clearly the direct search limit is not irrelevant.  If, for 
instance, the only information available were the direct search limit 
and the $A_{LR}$ measurement, we would conclude that the standard 
model is excluded at  90\% CL.  Theorists 
would have flooded the Los Alamos server with papers on the death of 
the standard model and the birth of new theories W,X,Y,Z...  In the 
SM fit the $A_{LR}$ measurement causes $m_{H}$ to 
shift by a factor two, from 170 to 85 GeV, and the 95\%
upper limit to fall from 570 to 260 GeV. It is 
fully weighted in the conventional standard model fit despite a 
significant contradiction with the standard model. 

If the discrepancy were even greater --- say, for instance, a 
precision measurement implying \footnote {In fact, parity violation in 
atomic Cesium currently implies $m_{H}\sim 11 $ MeV ({\it M}eV is not 
a typographical error) though only $1.2\sigma$ from 89 
GeV.\cite{deandrea} Its weight in the combined fit would be 
negligible.} $m_{H}=11$ MeV with a 99.9\% CL upper
limit at 89 GeV --- we would be faced with three alternatives: 1) omit 
the measurement from the SM fit, presuming a plausible reason exists 
to suspect a large systematic error, 2) disregard the search limits, 
presuming them to be systematically flawed in some way, or 3) to 
abandon the Standard Model.  On the other hand, a measurement one half 
standard deviation below the lower limit, with a $\simeq 30\%$
probability to be consistent with the limit, would surely be 
retained at essentially full weight.

The difficult question is how to resolve the intermediate cases in 
which the discrepancy is significant but not so significant that we 
are forced to choose between the data and the SM. Assuming the 
validity of the search limits and of the SM we consider a method that 
interpolates between the extremes of cases 1) and 2) above and which 
allows us to explore the sensitivity of the fit to the weight assigned 
to the discrepant measurements.  

The problem of how to combine inconsistent data has led to the 
break-up of many beautiful friendships.  The mathematical theory of 
statistics provides no magic bullets and ultimately the discrepancies 
can only be resolved by future experiments.  The PDG\cite{pdg} has for 
many years scaled the uncertainty of discrepant data sets by a factor
$$
S^{*} = \sqrt{\chi^{2}/(N-1)} \eqno(2)
$$
where $N$ is the number of measurements being combined.  They scale 
the uncertainty of the combined fit by the factor $S^{*}$ if and only 
if $S^{*} > 1$.  This is a conservative prescription, which amounts to 
requiring that the fit have a good confidence level, ranging from 32\% 
for $N=2$ to $\simeq 44\%$ for $N \simeq 10$.
If the confidence level is already good, the scale factor has 
little effect; it only has a major effect on very discrepant data.  
The PDG argues (see \cite{spdg}) that low confidence level fits occur 
historically at a rate significantly greater than expected by chance, 
that major discrepancies are often, with time, found to result from 
underestimated systematic effects, and that the scaled error provides 
a more cautious interpretation of the data.  

As an illustration we apply $S^{*}$ to the determination of $m_{H}$ 
from the nine measurements of $sin^{2}\theta^{lepton}_{eff}$.  The 
chi-squared for the nine measurements is 10.7 for 8 degrees of 
freedom, corresponding to CL = 0.20.  Then $S^{*}=\sqrt{10.7/8}=1.16$ 
and the conventional fit $sin^{2}\theta^{lepton}_{eff}=0.23148 \pm 
0.00021$ is modified to $0.23148 \pm 0.00024$.  The effect on $m_{H}$ 
is negligible:  the central value is unchanged, while the 95\%CL 
upper limit increases from 255 to just 272 GeV (using
\cite{ej} for $\alpha(m_{Z})$). The effect on $m_{H}$ is suppressed by 
the fact that the experimental error from $sin^{2}\theta^{lepton}_{eff}$ 
is dominated by the parametric error from $m_{t}$ and 
$\alpha(m_{Z})$ shown in table 1. Even for the more discrepant 
Summer 1997 data, with $\chi^{2}=14.6$ for 8 d.o.f. and CL = 0.07, 
the effect of the $S^{*}$ factor is moderate, with the 95\%CL 
upper limit increasing from 310 to 370 GeV.

We wish to construct an analogous method for situations in which the 
discrepancy is between some of a collection of measurements and a 
limit or physical boundary.  In analogy to the $\chi^{2}$ confidence 
level for $S^{*}$ our point of departure is the G-O-F CL 
(goodness-of-fit confidence level) between the measurements and the 
limit, for instance, equation (1) for the case at hand.  The method is 
to rescale the errors of the measurements that conflict with the limit 
by factors that increase the G-O-F CL of the rescaled data to a robust 
minimum value.  Following the PDG the minimum CL is chosen to equal 
the CL corresponding to $\chi^{2}=N-1$ for $N-1$ degrees of freedom.  
Regarding the limit as an additional degree of freedom we have $N=10$ 
for the nine measurements and the limit.  The minimum CL is then 0.44, 
corresponding to $\chi^{2}=9$ with 9 d.o.f.

Since there are two discrepant measurements, there are in general two 
different scale factors, $S_{\tau}$ and $S_{LR}$. In the notation of 
equation (1) the G-O-F CL  requirement is
$$
P_{9}(p'_{\tau},p'_{LR})= 0.44. \eqno{(3)}
$$
where 
$p'_{\tau}$ and $p'_{LR}$ are the values of $p_{\tau}$ and $p_{LR}$ 
after rescaling, 
$$
\delta (sin^{2}\theta^{lepton}_{eff})_{\tau} \rightarrow 
        S_{\tau}\cdot \delta (sin^{2}\theta^{lepton}_{eff})_{\tau} 
        \eqno{(4)}
$$ 
and 
$$   
       \delta (sin^{2}\theta^{lepton}_{eff})_{LR} \rightarrow 
	  S_{LR}\cdot \delta (sin^{2}\theta^{lepton}_{eff})_{LR}. 
	  \eqno{(5)}
$$
Equation (3) imposes one constraint, leaving a one dimensional parameter 
space within the ($S_{\tau}, S_{LR})$ plane to consider. 

Before turning to the electroweak 
data, we conclude this section with a general formulation of the method.
Consider a collection of $N$ measurements of a physical quantity $x$,
$$
\{x_{i},\delta_{i}\} \hspace{1in} i=1,\ldots,N  \eqno{(6)}
$$
where the $x_{i}$ are the individual measured values and $\delta_{i}$ 
are the one standard deviation uncertainties.  Suppose there is 
an exact lower limit or physical boundary (this assumption is 
relaxed in section 5 for the Higgs boson search limits),
$$
             x_{\rm TRUE} > x_{\rm LIMIT},  \eqno{(7)}
$$
and that  $n \leq N$ of the measurements fall below the limit,
$$
              x_{i} < x_{\rm LIMIT} \hspace{1in} i=1,\ldots,n  
$$
$$
              x_{i} > x_{\rm LIMIT} \hspace{1in} i=n+1,\ldots,N.  
              \eqno{(8)}
$$
Furthermore assume, in analogy to $p_{\tau}$ and $p_{LR}$ defined 
above, that the probability density function associated with each of 
the $n$ low measurements, $PDF_{i}(x-x_{i},\delta_{i})$ implies a 
probability $p_{i}$ that the measurement conflicts with the limit (7),
$$
    p_{i}=\int^{x_{\rm LIMIT}}_{-\infty} PDF_{i}(x-x_{i},\delta_{i})\ 
    dx. 
    \eqno{(9)} 
$$

By analogy with equation (1) we compute an upper bound on the G-O-F CL 
between the $N$ measurements and the limit. We order the $n$ low 
measurements such that $p_{1}>p_{2}>\ldots>p_{n}$. The upper bound is 
then obtained by assuming 
$$
x_{\rm TRUE}=x_{\rm LIMIT}   \eqno{(10)}
$$ 
and computing the probability that {\em any} $n$ of the $N$ 
measurements, designated by ordered integer $n$-tuples $\{a_{1},\ldots 
,a_{n}\}$ chosen from the integers ${\{1,\ldots ,N}\}$, ordered such 
that $p_{a_{1}}>p_{a_{2}}>\ldots>p_{a_{n}}$, satisfy the 
condition
$$
p_{a_{i}}\geq p_{i}       \eqno{(11)}
$$
for all $i=1,\ldots ,n$. 

The combined PDF for the $N$ independent measurements is
$$
PDF_{N}(\{x-x_{i},\delta_{i}\})= \prod_{i=1}^{N}
                                 PDF_{i}(x-x_{i},\delta_{i}).
                                 \eqno{(12)}
$$ 
Finally we can write the upper bound on the G-O-F CL between the $N$ 
measurements and the limit in the general form
$$
	 P_{N}(p_{1},\ldots .p_{n})= \sum_{\{a_{1},\ldots ,a_{n}\}} 
	 \int_{D} PDF_{N}(\{x_{a_{i}}-x_{\rm LIMIT},\delta_{a_{i}}\})
	 \ dx_{a_{1}}\ldots dx_{a_{n}}
	 \eqno{(13)}
$$
where the sum is over all ordered integer $n$-tuples $\{a_{1},\ldots 
,a_{n}\}$ chosen from the integers ${\{1,\ldots ,N}\}$ and 
the domain of integration $D$ is defined by the condition 
$$
{x_{\rm LIMIT}-x_{a_{i}}\over \delta_{a_{i}} } \geq 
	        {x_{LIMIT}-x_{i} \over \delta_{i} }     \eqno{(14)} 
$$
for all $i=1,\ldots ,n$.  

Equations (12-14), in all their obtuse generality, are just the 
straightforward generalization of the G-O-F CL $P_{9}(p_{1},p_{2})$ 
given explicitly in equation (1).  The general statement of the method 
now closely follows that example.  We require a minimum G-O-F CL
$$
P_{N}(p_{1},\ldots .p_{n}) \geq P_{\rm MIN}      \eqno{(15)}
$$
where $P_{\rm MIN}$ is the confidence level corresponding to the 
chi-squared distribution with $\chi^{2}=N$ for $N$ degrees of 
freedom. If equation (15) is satisfied by the data we  combine the 
data without further ado. If equation (15) is not obeyed we rescale 
the errors of the $n$ low measurements,
$$
\delta_{i} \rightarrow \delta'_{i} = S_{i}\delta_{i}  \eqno{(16)},
$$
so that the $p_{i}$ defined in equation (9) are replaced by $p'_{i}$ 
$$
	p'_{i}=\int^{x_{\rm LIMIT}}_{-\infty} 
	PDF_{i}(x-x_{i},\delta'_{i})\ dx.  \eqno{(17)}
$$
such that the G-O-F CL for the scaled data satisfies the requirement, 
$$
P_{N}(p'_{1},\ldots .p'_{n}) =P_{\rm MIN}.      \eqno{(18)}
$$
The condition equation (18) is satisfied by an $n-1$ dimensional 
subspace of the space of $n$-tuples $(S_{1},\ldots ,S_{n})$.

This section concludes with a brief discussion of the relationship of 
the scale factor method to the Cousins-Feldman definition of 
confidence intervals near a physical boundary.\cite{cf} They observe 
that the standard construction of confidence intervals near a physical 
boundary is flawed, in that it leads to intervals that in some 
instances ``under-cover'' (i.e., correspond to less than the nominal 
probability) and which have discontinuities as a function of the 
central value that are artifacts of the construction.  
Particularly germane to the method presented here is their 
observation that near a boundary the conventional construction 
confuses two aspects of the fit that are or should be conceptually 
distinct: that is, the goodness-of-fit CL between the measurement and 
the limit is typically assessed based on the extent that the 
conventional confidence intervals obtained from the fit overlap the 
region allowed by the boundary or limit.  In contrast, the usual 
procedure for combining data (away from a boundary) uses the minimum 
of the chi-squared distribution to asses goodness-of-fit, while the 
confidence intervals are obtained quite independently from the shape 
of the chi-squared distribution.

They propose confidence intervals which rectify these shortcomings, at 
the cost of relaxing the upper limits near the boundary.  In 
particular, their confidence intervals only have support in the 
allowed region, leaving the assessment of goodness-of-fit as a 
separate issue.  In this paper I use a goodness-of-fit estimator, 
$P_{N}(p_{1},\ldots,p_{n})$, which is quite distinct from the 
confidence intervals that are the output of the fit.  Rather the 
goodness-of-fit estimator is computed at the outset and is then used 
to constrain the scale factors that determine the final fit and 
confidence intervals.  The upper limits on $m_{H}$ obtained from the 
scaled fits are given with the Cousins-Feldman construction, though 
for comparison the conventionally defined limits are also provided.

\noindent{\bf {4. Scaled standard model fits}}

In this section the scale factor method is applied to the SM Higgs 
boson mass fit.  We indicate how the scaled fit is obtained and 
present the results.  The results in this section are obtained under 
the assumption that the search limit is perfect, i.e., $m_{H}>89.3$ 
GeV at 100\%CL. In section 5 I show that
essentially the same results follow from the actual data of the search 
experiments, as a result of the rapidly rising confidence level for 
exclusion limits below 89.3 GeV.

The results are shown in tables 5 and 6 and in figure 1. Consider for 
instance the results using the more conservative evaluation\cite{ej} 
of $\alpha(m_{Z})$, shown in table 5 and in the solid curves in 
figure 1.  Recall from section 2 that the goodness-of-fit CL between 
the nine measurements and a perfect lower limit at 89.3 GeV is 12\%.  
Table 5 displays a selection of scaled fits with G-O-F CL of 44\%.
At one extreme the $A_{LR}$ measurement is unscaled, $S_{LR}=1$, 
while $S_{\tau}=3.5$. The effect on the SM fit is negligible: the 
central value and 95\%CL upper limits for $m_{H}$ increase by just 
$\simeq 15\%$. At the other extreme, if we attempt to leave 
$A_{FB}^{\tau}$ unscaled, $S_{\tau}=1$, we find that even if  
$A_{LR}$ is removed from the fit, $S_{LR} \rightarrow \infty$, the 
G-O-F CL is 39\%. At this extreme in order to reach 44\% it is 
necessary to set $S_{\tau}=1.06$ and $S_{LR} \rightarrow \infty$. The 
effect on the fit is maximal: the central value increases to 
$m_{H}=175$ GeV and the 95\%CL upper limit increases to 750 GeV.

The scaled fits are obtained numerically, as described below.  
Consider for instance the entry in table 5 with $S_{LR}=1$.  From 
table 3 we see that $p'_{LR}=p_{LR}=0.932$.  Equation (1), 
$P_{9}(p'_{\tau},0.932)=0.44$, is solved numerically to obtain 
$p'_{\tau}=0.684$.  Assuming Gaussian statistics we then deduce from 
the Gaussian distribution that $sin^{2}\theta^{lepton}_{eff}$ from 
$A_{FB}^{\tau}$ lies $0.475\delta_{\tau}^{\prime\ {\rm TOTAL}}$ below 
$sin^{2}\theta^{lepton}_{eff}=0.23151$, the latter being the value of 
$sin^{2}\theta^{lepton}_{eff}$ corresponding to $m_{H}^{\rm 
LIMIT}=89.3$ GeV. Here `TOTAL' in $\delta_{\tau}^{\prime\ {\rm 
TOTAL}}$ denotes the sum in quadrature of the rescaled experimental 
error $\delta'_{\tau}$ and the parametric error from the sources shown 
in table 1,
$$
\delta_{\tau}^{'\ {\rm TOTAL}}=\sqrt{\delta_{\tau}^{'2} + 
\delta_{P}^{2}}. 
   \eqno{(19)}
$$
Taking $sin^{2}\theta^{lepton}_{eff}=0.22987$ from $A_{FB}^{\tau}$ we 
then obtain $\delta_{\tau}^{\prime\ {\rm TOTAL}}= (0.23151 - 
0.22987)/0.475 = 0.00345$.  Using reference \cite{ej} the effective 
parametric error, expressed as an equivalent uncertainty in 
$sin^{2}\theta^{lepton}_{eff}$ is 0.00028, so that\footnote{The 
parametric error is negligible compared to $\delta'_{\tau}$ but is 
important relative to more precise measurements such as 
$\delta_{LR}$.} $ \delta'_{\tau} = 0.00344$, from which we finally 
obtain $S_{\tau}= \delta'_{\tau}/\delta_{\tau}= 0.00344/000.98=3.51$.

The fits for the intermediate cases are obtained similarly, by 
fixing either $S_{\tau}$ or $S_{LR}$ and computing the other.  
Equivalently, one may choose a grid in $p'_{\tau}$ or $p'_{LR}$ and 
compute the other, from which all other quantities in the fit can 
be obtained.  (The latter was the procedure actually followed 
to construct tables 5 and 6). 

Except for a small ``central plateau'' it is clear from the
tables and figure that the value of $m_{H}$ is dominated by 
$S_{LR}$, as expected from the importance of $A_{LR}$ in the fit.  
In table 5 the ``central plateau'' occurs between $S_{LR}=1.75$ and 
$S_{LR}=2.01$, for which the inverse effects of increasing $S_{LR}$, 
and decreasing $S_{\tau}$ cancel one another.  At the extreme of table 
5, with $S_{LR} \rightarrow \infty$, the value of 
$sin^{2}\theta^{lepton}_{eff}$ is greater than the conventional fit 
value by two standard deviations, while the central value of $m_{H}$ 
is increased by one standard deviation.  The shift in $m_{H}$ is 
smaller than the shift in $sin^{2}\theta^{lepton}_{eff}$ because of 
the diluting effect of the parametric error in table 1.

Table 6 and the dashed lines in figure 1 are based on $\alpha(m_{Z})$ 
from reference \cite{dh98}.  They display the same general features as 
the fits based on \cite{ej}.  The central values for $m_{H}$ are 
larger while the 95\% CL upper limits are smaller, because reference 
\cite{dh98} finds larger $\alpha(m_{Z})$ but with smaller claimed 
uncertainty, and the latter effect dominates the former in the 
determination of the upper limit.  Because the central values are 
larger, the discrepancies with the search limits are somewhat reduced 
(cf.  tables 3 and 4) and consequently the scale factors are smaller.  
In the extreme case it is possible to satisfy the G-O-F CL requirement 
of 44\% for $S_{\tau}=1$ and finite $S_{LR}$.  The fit in that case, 
with $S_{LR}=3.6$, yields $m_{H}=207$ GeV and $m_{H}<670$ GeV. at 
95\% CL. 

\noindent{\bf {5. Including the search limit confidence levels}}

In the previous sections we regarded the search limit, 
$m_{H}>89.3$ GeV, as an absolute boundary, neglecting the fact that 
it carries a less than perfect 95\% confidence level. In this section 
we will see that the finite  confidence level has negligible effect 
on the scaled fits and that the results presented in section 4 apply 
to the actual experimental situation.

The conclusion follows from the rather steep dependence of the Higgs 
boson search limit confidence level as a function of $m_{H}^{\rm 
LIMIT}$.  For instance, preliminary data\cite{alephlimits} from the 
ALEPH experiment show that the confidence level for 
$m_{H}>m_{H}^{\rm LIMIT}$ is 95\% at
$m_{H}^{\rm LIMIT}=88$ GeV, rising to 99\% at 83 GeV and to 99.9\% at 
78 GeV. These values are conservative since they follow from just one 
of the four LEP experiments.  Furthermore the conclusion reached below 
that the results of section 4 apply to the real experimental limits 
does not depend at all sensitively on the values quoted above for 
$m_{H}^{\rm LIMIT}$ at 99\% and 99.9\%, since the dependence on 
$m_{H}^{\rm LIMIT}$ is logarithmic.

To get an upper limit on the correction to the ``perfect search 
limit'' results of section 4 we consider fits using the evaluation of 
$\alpha(M_{Z})$ claiming greater precision\cite{dh98}, since those fits 
are most sensitive to the value of $m_{H}^{\rm LIMIT}$.  Consider the 
goodness-of-fit CL for the unscaled data.  We refine the notation, 
making explicit the dependence of the probabilities $p_{i}$ defined in 
equation (9) on $m_{H}^{\rm LIMIT}$, by writing $p_{\tau}(m_{H}^{\rm 
LIMIT})$ and $p_{LR}(m_{H}^{\rm LIMIT})$.  Notice from equation (9) 
that these probabilities are defined for perfect search limits.  The 
actual goodness-of-fit CL can be obtained by weighting the value 
for a perfect search limit at 89.3 GeV by its actual 95\% CL, i.e.,
$0.95\cdot 0.181$, and then integrating over the corresponding 
larger goodness-of-fit CL's for smaller values of $m_{H}^{\rm 
LIMIT}$, $P_{9}(p_{\tau}(m_{H}^{\rm LIMIT}),p_{LR})(m_{H}^{\rm 
LIMIT}))$, weighted by the probability measure given by the 
derivative of the experimental search limit confidence level with 
respect to $m_{H}^{\rm LIMIT}$.

In practice it suffices to obtain an upper limit by approximating the 
integral by a discrete sum over a few regions, representing the 
goodness-of-fit CL for each region by the maximum for the region, 
which occurs at the lower boundary of the region in $m_{H}^{\rm 
LIMIT}$.  In the present instance just two regions will suffice, 
corresponding to the 99 and 99.9\% limits quoted above.  To an
accuracy of $\pm 0.001$ the upper limit on the true 
goodness-of-fit CL is given by
$$
P_{9}^{\rm COMBINED}= 0.95P_{9}\left (p_{\tau}(89.3\ {\rm GeV}),
                       p_{LR}(89.3\ {\rm GeV})\right ) 
$$
$$
			 +\; 0.049P_{9}\left (p_{\tau}(83\ {\rm GeV}),p_{LR}(83\ 
			 {\rm GeV})\right )
$$
$$
			 +\; 0.001P_{9}\left (p_{\tau}(78\ {\rm GeV}),p_{LR}(78\ 
			 {\rm GeV})\right ).  \eqno{(20)}
$$
The relevant values of $p_{\tau}$, $p_{LR}$ and $P_{9}$ are given in 
table 7.  Substituting those values into equation (20) we find that 
the actual G-O-F CL is bounded above by 0.183 with an uncertainty 
$\pm 0.001$.  This value differs hardly at all from the 0.181 CL 
that corresponds to a perfect search limit at 89.3 GeV.

Since the scaled data is less precise, the correction due to the 
actual confidence limits of the searches will be even smaller and is 
therefore also perfectly negligible for the scaled fits.  (I have 
verified this by applying the above analysis to some of the scaled 
fits, including the most sensitive case, from table 6 with 
$S_{LR}=1$.)  In fact, the numerical error in calculating tables 5 and 
6 is of order 0.01, much bigger than the 0.002 correction from the 
finite confidence level of the search limits.  We conclude that the 
fits shown in tables 5 and 6 do in fact reflect  the actual 
experimental confidence levels of the direct search limits. 

\noindent{\bf {6. Conclusion}}

Motivated by the observation that within the SM framework two of the 
nine measurements of $sin^{2}\theta^{lepton}_{eff}$ are individually 
in significant conflict with the SM Higgs boson direct search limit, 
we constructed a scale factor method based on an aggregate 
goodness-of-fit confidence level between the complete set of nine 
measurements and the limit.  Like an analogous scale factor used for 
many years by the Particle Data Group, the scale factor proposed here 
is intended to account for the possibility of underestimated 
systematic effects.  It is applicable to other physical situations in 
which some of a set of measurements are in conflict with a physical 
boundary or experimental limit.  Applied to the SM Higgs boson mass, 
the scaled fits exhibit the dependence of the fit on the 
weight accorded to the two measurements that are in conflict with the 
search limits.  The fits in which the weight of $A_{LR}$ is reduced 
allow a central value of $m_{H}$ as large as $\simeq 200$ GeV and a 
95\%CL upper limit as
large as 750 GeV. Relative to the conventional least-square fit, the 
central value of $sin^{2}\theta^{lepton}_{eff}$, increases by as 
much as two standard deviations while $m_{H}$ increases by as much 
as one standard deviation.

There is a tendency to think that the value of 
$sin^{2}\theta^{lepton}_{eff}$ is only of interest as a prognosticator 
of the Higgs boson mass, so that it will be of only secondary interest 
after/if a Higgs boson is discovered.  This view underestimates the 
importance of $sin^{2}\theta^{lepton}_{eff}$ as a fundamental probe of 
a variety of new physics, not simply restricted to the Higgs sector.  
By comparing the measured value of $sin^{2}\theta^{lepton}_{eff}$ with 
the value predicted by the directly measured mass of the Higgs boson, 
we would have a probe of other possible new physics, such as for 
instance extended gauge sectors or nonsinglet heavy quanta.  It would 
therefore be regrettable if the brilliant program of precision studies 
of $Z$ particle properties were to conclude with some measure of 
uncertainty as to how definitively the value of 
$sin^{2}\theta^{lepton}_{eff}$ has been determined.

There are a variety of possible explanations for the anomalies that 
have affected the measurements of $sin^{2}\theta^{lepton}_{eff}$, both 
the internal inconsistencies, which have diminished but continue to 
exist as of this writing, and the inconsistencies with the search 
limits that are the subject of this paper.  They may in fact simply be 
the result of bad luck, chance fluctuations.  They may result from 
underestimated systematic errors among some of the measurements.  Or 
they may represent real effects and be harbingers of new physics.  
Hopefully the situation will be clarified by further experimental 
work, beginning with new data and/or analyses to be presented at the 
Summer 1998 conferences.

\noindent{\bf Acknowledgements:} I wish to thank Mark Strovink for 
constructive criticism of the analysis presented in reference 
\cite{msc}.  This work was supported by the Director, Office of Energy 
Research, Office of High Energy and Nuclear Physics, Division of High 
Energy Physics of the U.S. Department of Energy under Contract 
DE-AC03-76SF00098.

\newpage
\vskip 0.5in
\begin{center}
\noindent {\bf Tables}
\end{center}
\vskip 0.2in

\noindent Table 1.  Uncertainties in the evaluation of the natural 
logarithm of the SM Higgs boson mass, ${\rm ln}\ m_{H}$, from 
$sin^{2}\theta^{lepton}_{eff}$.  The two values for $\alpha(m_{Z})$ 
and `Total' correspond to references \cite{ej} (larger values) and 
\cite{dh98} (smaller values).

\begin{center}
\vskip 20pt

\begin{tabular}{cc}
\hline
\hline 
 Parameter        & $\Delta ({\rm ln}\ m_{H})$ \\
\hline 
$sin^{2}\theta^{lepton}_{eff}$ & 0.40 \\
$\alpha (m_{Z})$ & 0.46 or 0.11 \\         
$m_{t}$ &  0.32 \\
$\alpha_{S} (m_{Z})$ & 0.02 \\
theory  &  0.07 \\
\hline
Total & 0.67 or 0.52 \\
\hline
\hline
\end{tabular}
\end{center}

\vskip 1in

\noindent Table 2.  Individual measurements of 
$sin^{2}\theta^{lepton}_{eff}$ with $1\sigma$ experimental errors and 
their pulls with respect to the least-square fit value 
$sin^{2}\theta^{lepton}_{eff}=0.23148 \pm 0.00021$, listed in the 
order of the absolute value of the pulls.

\begin{center}
\vskip 20pt
\begin{tabular}{ccc}
\hline
\hline 
Measurement & $sin^{2}\theta^{lepton}_{eff}$ & Pull \\
\hline
$A_{FB}^{\tau}$ & 0.22987 (98) & -1.61 \\
$A_{LR}$ & 0.23084 (35) & -1.57 \\
$A_{FB}^{b}$ & 0.23211 (39) & +1.42 \\
$A_{\tau}$ & 0.23241 (80) & +1.12 \\
$<Q_{FB}>$ & 0.23210 (100) & +0.60 \\
$A_{e}$ & 0.23193 (90) & +0.48 \\
$A_{FB}^{c}$ & 0.23160 (110) & +0.12 \\
$A_{FB}^{e}$ & 0.23164 (145) & +0.11 \\
$A_{FB}^{\mu}$ & 0.23147 (82) & +0.01 \\
\hline
\hline
\end{tabular}
\end{center}

\newpage
\vskip 1in

\noindent Table 3.  SM Higgs boson mass prediction for the individual 
measurements, based on $\alpha(m_{Z})$ from reference \cite{ej}, 
listed in order of the precision of the measurements.  The central 
value of $m_{H}$ is shown along with the symmetric (in ln\ $m_{H}$) 
90\% confidence interval $m_{95}^{>},m_{95}^{<}$
and the implied probability that $m_{H}<89.3$ GeV.

\begin{center}
\vskip 20pt
\begin{tabular}{cccc}
\hline
\hline 
Measurement & $m_{H}$ (GeV) & $m_{95}^{>},m_{95}^{<}$ & 
$P(m_{H}<89.3\ {\rm GeV})$ \\
\hline
$A_{LR}$ & 25 & 6, 100 & 0.93\\
$A_{FB}^{b}$ & 280 & 62, 1300 & 0.11 \\
$A_{\tau}$ & 500 & 35, 7100 & 0.14 \\
$A_{FB}^{\mu}$ & 83 & 5, 1300 & 0.52 \\
$A_{e}$ & 200 & 10, 3800 & 0.33\\
$A_{FB}^{\tau}$ & 4 & 0.2, 95 & 0.95 \\
$Q_{FB}$ & 280 & 11, 7200 & 0.29 \\
$A_{FB}^{c}$ & 110 & 4, 2800 & 0.47 \\
$A_{FB}^{e}$ & 110 & 1, 12000 & 0.47 \\
\hline
\hline
\end{tabular}
\end{center}

\vskip 1in

\noindent Table 4.  Same as table 3 but with $\alpha(m_{Z})$ from reference 
\cite{dh98}.

\begin{center}
\vskip 20pt
\begin{tabular}{cccc}
\hline
\hline 
Measurement & $m_{H}$ (GeV) & $m_{95}^{>},m_{95}^{<}$ & 
$P(m_{H}<89.3\ {\rm GeV})$ \\
\hline
$A_{LR}$ & 33 & 10, 110 & 0.91\\
$A_{FB}^{b}$ & 370 & 100, 1400 & 0.04 \\
$A_{\tau}$ & 660 & 50, 8600 & 0.10 \\
$A_{FB}^{\mu}$ & 110 & 8, 1500 & 0.45 \\
$A_{e}$ & 260 & 15, 4700 & 0.27\\
$A_{FB}^{\tau}$ & 5 & 0.2, 120 & 0.93 \\
$Q_{FB}$ & 360 & 15, 8800 & 0.24 \\
$A_{FB}^{c}$ & 140 & 6, 3400 & 0.41 \\
$A_{FB}^{e}$ & 150 & 2, 15000 & 0.42\\
\hline
\hline
\end{tabular}
\end{center}

\newpage
\vskip 1in

\noindent Table 5. Fits based on $\alpha(m_{Z})$ from reference 
\cite{ej}. The first line is the conventional fit while the other 
lines display scaled fits that meet the 44\% minimum goodness-of-fit 
confidence level for the measurements and search limit.  For each fit, 
specified by the pair of scale factors $S_{\tau}, S_{LR}$, the table 
displays the fitted value of $sin^{2}\theta^{lepton}_{eff}$ with 
$1\sigma$ uncertainty, the central value of $m_{H}$, the conventional 
95\% CL upper limit, $m_{95}$, and the Cousins-Feldman\cite{cf} 95\% 
CL upper limit, $m_{95}^{\rm CF}$. 

\begin{center}
\vskip 20pt
\begin{tabular}{cccccc}
\hline
\hline 
$S_{LR}$ & $S_{\tau}$ & $sin^{2}\theta^{lepton}_{eff}$ & $m_{H}$ & 
$m_{95}$ & $m_{95}^{\rm CF}$\\
\hline 
1 & 1 & 0.23148 (21) & 85 & 260 & 320 \\
1 & 3.51 & 0.23155 (22) & 97 & 300 & 370 \\
1.11 & 2.27 & 0.23160 (22) & 105 & 320 & 400 \\
1.26 & 1.87 & 0.23165 (23) & 117 & 370 & 460 \\
1.42 & 1.74 & 0.23170 (24) & 127 & 410 & 510 \\
1.59 & 1.71 & 0.23173 (24) & 137 & 440 & 550 \\
1.78 & 1.68 & 0.23177 (25) & 146 & 480 & 600 \\
2.01 & 1.28 & 0.23177 (25) & 147 & 480 & 600 \\
2.50 & 1.16 & 0.23180 (26) & 154 & 510 & 640 \\
$\infty$ & 1.06 & 0.23186 (27) & 175 & 590 & 750 \\
\hline
\hline
\end{tabular}
\end{center}

\newpage 
\vskip 1in

\noindent Table 6. As in table 5 but with $\alpha(m_{Z})$ from reference 
\cite{dh98}. 

\begin{center}
\vskip 20pt
\begin{tabular}{cccccc}
\hline
\hline 
$S_{LR}$ & $S_{\tau}$ & $sin^{2}\theta^{lepton}_{eff}$ & $m_{H}$  & 
$m_{95}$ & $m_{95}^{\rm CF}$\\
\hline 
1 & 1 & 0.23148 (21) & 112 & 260 & 310 \\
1 & 1.84 & 0.23154 (21) & 124& 295& 350 \\
1.07 & 1.71 & 0.23157 (22) & 131 & 310 & 370 \\
1.18 & 1.60 & 0.23161 (23) & 143& 350 & 420 \\
1.31 & 1.57 & 0.23165 (23) & 155 & 385 & 460 \\
1.45 & 1.54 & 0.23169 (24) & 167 & 420 & 500 \\
1.62 & 1.18 & 0.23170 (24) & 169 & 430 & 520 \\
1.75 & 1.12 & 0.23171(24) & 173 & 440 & 530 \\
2.00 & 1.08 & 0.23174 (25) & 182 & 470 & 570 \\
3.62 & 1.00 & 0.23181 (26) & 207 & 550 & 670 \\
\hline
\hline
\end{tabular}
\end{center}

\vskip 1in

\noindent Table 7.  The goodness-of-fit confidence level between the 
nine $sin^{2}\theta^{lepton}_{eff}$ measurements and the direct search 
limit for $m_{H}^{\rm LIMIT}$ corresponding to experimental confidence 
levels of 95\%, 99\%, and 99.9\%.  The G-O-F
CL's, $P_{9}(p_{\tau}, p_{LR}$), are computed assuming perfect search 
limits at each $m_{H}^{\rm LIMIT}$, as discussed in the 
text.  Reference \cite{dh98} is used for $\alpha(m_{Z})$.

\begin{center}
\vskip 20pt
\begin{tabular}{ccccc}
\hline
\hline 
Search Limit CL & $m_{H}^{\rm LIMIT}({\rm GeV})$ & $p_{\tau}$ & 
$p_{LR}$ & $P_{9}(p_{\tau}, p_{LR}$)\\
\hline
95\% & 89.3 & 0.933 & 0.910 & 0.181 \\
99\% & 83    & 0.928 & 0.894 & 0.225 \\
99.9\% & 78 & 0.924 & 0.878 & 0.264 \\
\hline
\hline
\end{tabular}
\end{center}

\newpage
\begin{center}
{\bf Figure Caption}
\end{center}
\vskip .5 in

\noindent Figure 1.  Scaled fits that meet the minimum goodness-of-fit 
criterion.  The central value and 95\%CL upper limit
for the Higgs boson mass are plotted as a function of the scale factor 
for $sin^{2}\theta^{lepton}_{eff}$ from $A_{LR}$.  Solid and dashed 
lines correspond to the evaluations of $\alpha(m_{Z})$ from references 
\cite{ej} and \cite{dh98} respectively.

\begin{figure}
\epsfbox[72 72 300 600]{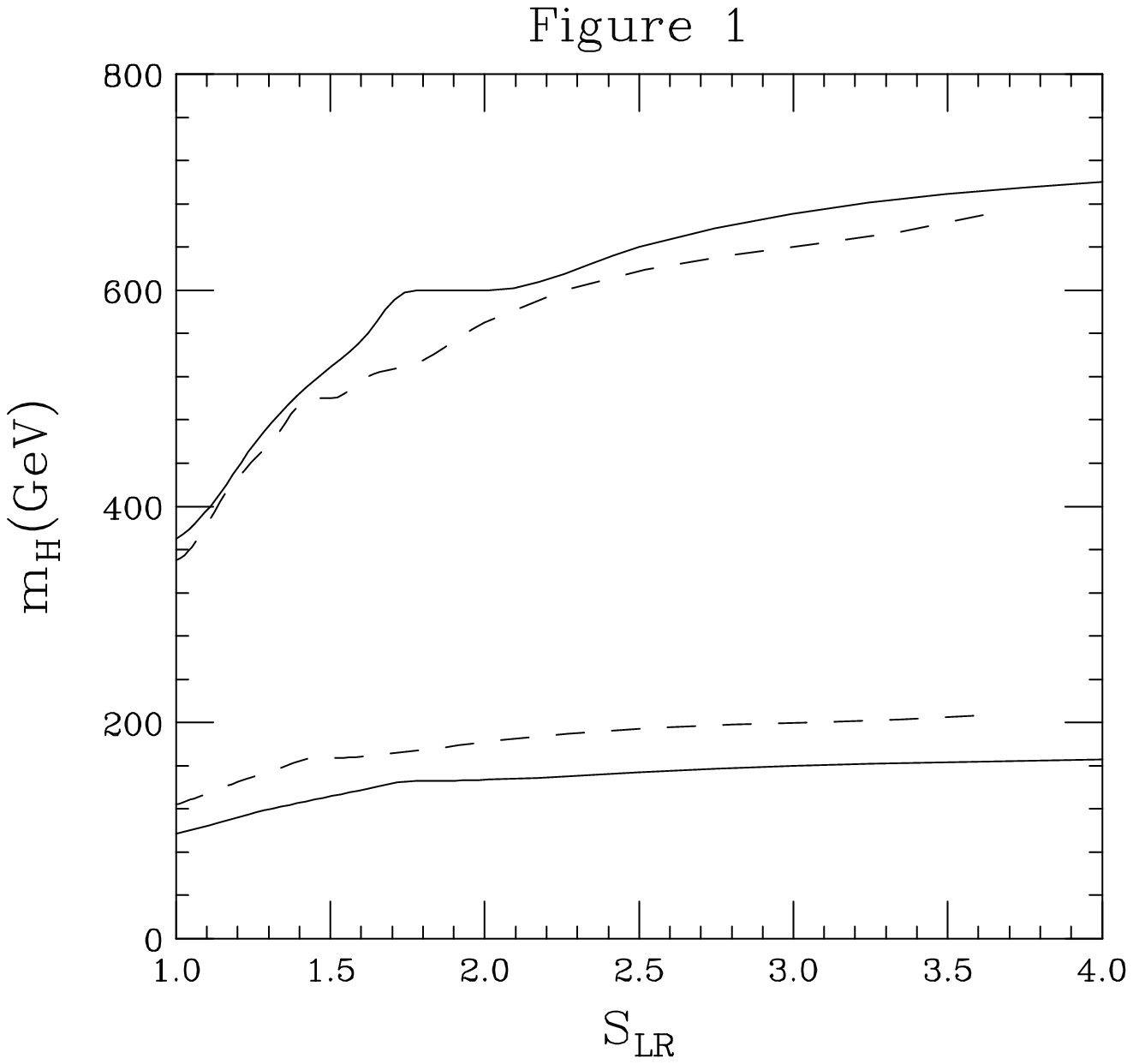}
\end{figure}

\end{document}